\documentclass[sigconf]{acmart}
\usepackage{xr}
\usepackage[T1]{fontenc}
\usepackage{nameref}
\usepackage{flushend}
\usepackage{mdframed}
\usepackage{multirow,slashbox,pbox,booktabs}
\usepackage[export]{adjustbox}
\usepackage{amsmath,amsfonts,amssymb,amsthm,paralist}
\usepackage{algorithmicx}
\usepackage[normalem]{ulem}
\usepackage[noend]{algpseudocode}
\usepackage[inline]{enumitem}

\makeatletter \let\c@table\c@figure \makeatother

\graphicspath{{../}}

\def\etal{\textit{et al.}}

\copyrightyear{2017}
\acmYear{2017}
\setcopyright{acmcopyright}
\acmConference{KDD'17}{}{August 13--17, 2017, Halifax, NS, Canada.}
\acmPrice{15.00}
\acmDOI{http://dx.doi.org/10.1145/3097983.3098146}
\acmISBN{978-1-4503-4887-4/17/08}

\title{Relay-Linking Models for Prominence\\ and Obsolescence in Evolving Networks}

\author{Mayank Singh}
\affiliation{%
  \institution{Dept.  of Computer Science and Engg.}
  \streetaddress{IIT Kharagpur, India}
  }
\email{mayank.singh@cse.iitkgp.ernet.in}

\author{Rajdeep Sarkar}
\affiliation{%
  \institution{Department of Mathematics}
 \streetaddress{IIT Kharagpur, India}
  }
\email{rajdeep.sarkar@iitkgp.ac.in}

\author{Pawan Goyal}
\affiliation{%
  \institution{Dept.  of Computer Science and Engg.}
  \streetaddress{IIT Kharagpur, India}
  }
\email{pawang@cse.iitkgp.ernet.in}

\author{Animesh Mukherjee}
\affiliation{%
  \institution{Dept.  of Computer Science and Engg.}
  \streetaddress{IIT Kharagpur, India}
  }
 \email{animeshm@cse.iitkgp.ernet.in}

\author{Soumen Chakrabarti}
\affiliation{%
  \institution{Dept.  of Computer Science and Engg.}
 \streetaddress{IIT Bombay, India}
  }
\email{soumen@cse.iitb.ac.in}

\renewcommand{\shortauthors}{Singh et al.}

\ccsdesc[500]{Computing methodologies~Modeling methodologies}
\vspace{-0.3cm}
\keywords{Network growth models, relay-link, aging models}
\begin{document}

\begin{abstract}  
The rate at which nodes in evolving social networks acquire links
(friends, citations) shows complex temporal dynamics.  
Preferential attachment and link copying models, while enabling elegant analysis, only capture rich-gets-richer effects, not aging and decline.  
Recent aging
models are complex and heavily parameterized; most involve estimating
1--3 parameters \emph{per node}.  These parameters are
\emph{intrinsic}: they explain decline in terms of
events in the past of the same node, and do not explain, \emph{using
  the network}, where the linking attention might go instead.  We
argue that traditional characterization of
linking dynamics are insufficient to judge the faithfulness of models.
We propose a new \emph{temporal sketch} of an evolving graph, and
introduce several new characterizations of a network's temporal
dynamics.  Then we propose a new family of frugal aging models with no
per-node parameters and only two global parameters.  Our model is based on
a surprising \emph{inversion} or undoing of triangle completion, where
an old node \emph{relays} a citation to a younger follower in its
immediate vicinity.  Despite very few parameters, the new family of models shows remarkably better fit with 
real data. Before concluding, we analyze temporal signatures for  various  research  communities  yielding 
further  insights  into their  comparative  dynamics. To facilitate reproducible research, we shall soon make
all the codes and the processed dataset available in the public domain.
\end{abstract}

\maketitle

\section{Introduction}
How do actors in a social network pass from prominence to
obsolescence and obscurity?  Is aging intrinsic, or informed and
influenced by the local network around actors?  And how does the aging
process affect properties of social networks, specifically, the
tension between entrenchment of prominence (aka ``rich gets richer''
or the Matthew effect) vs.\ obsolescence?  These are fundamental
questions for any evolving social network, but particularly
well-motivated in bibliometry.
With rapidly growing publication repositories, understanding the
networked process of obsolescence is as important to the emerging field
of \emph{academic
analytics}\footnote{\protect\url{https://en.wikipedia.org/wiki/Academic_analytics}}
as understanding the rise to prominence.

In his classical papers, Price~\cite{deSollaPrice510,ASI:ASI4630270505} presents evidences of obsolescence in bibliography network. 
Recently, Parolo \etal\ \cite{Parolo2015734} presented evidence that it is indeed
becoming ``increasingly difficult for researchers to keep track of all
the publications relevant to their work'', which can lead to
reinventions, redundancies, and missed opportunities to connect ideas.
Based on analysis of citation data, they propose a pattern of a
paper's citation counts per year, which peaks within a few years and
then the typical paper fades into obscurity.  Such works have seen
considerable press following, with headlines\footnote{http://www.independent.co.uk/news/science/there-are-too-many-studies-new-study-finds-10101130.html} ranging from the
tongue-in-cheek ``Study shows there are too many studies" to the more
alarmist ``Science is `in decay' because there are too many studies''.

On the other hand,
Verstak \etal\ \cite{DBLP:journals/corr/VerstakASHILS14} claim that
fear of evanescence is misplaced, and that older papers account for an
increasing fraction of citations as time passes.  In a related vein,
when PageRank began to be used for ranking in Web search, there was a
concern that older pages have an inherent --- and potentially unfair
--- advantage over emerging pages of high quality, because they have
had more time to acquire hyperlink citations.  In fact, algorithms
have been proposed to compensate for this
effect \cite{cho2005page,pandey2005shuffling}.  (In that domain,
clickthrough also provides valuable support for recency to combat
historic popularity.)

So where does reality lie between entrenchment and obsolescence?
Chakraborty \etal\ \cite{Chakraborty:2015:CSC:2817191.2701412} present
a nuanced analysis that naturally clusters papers into the ephemeral
and the enduring.  This gives hope that not all creativity is lost in
the sands of time; but neither do older papers capture all our
attention.  Others \cite{wang2013quantifying,Wang20094273} model aging
as intrinsic to a paper, reducing the probability of citing it as it
ages, but do not prescribe where the diverted citations end up.

In an interesting work on explaining aging by attention stealing, Waumans \etal\ \cite{waumans2016genealogical} present several evidences of attention stealing from parent paper by child paper.
They show that the arXiv\footnote{https://arxiv.org/} article titled ``Notes on D-Branes"~\cite{polchinski1996notes} published in the year 1996 started losing its citations in the very next year (1997). The reason for attention stealing is attributed to four papers that cite~\cite{polchinski1996notes} and go further on the same topic.
In another example, the paper titled ``Theory of Bose-Einstein condensation in trapped gases'' ~\cite{dalfovo1999theory} from the American Physical Society dataset\footnote{http://journals.aps.org/datasets} suffers from a similar stealing effect. This paper starts losing attention to its three child papers six years after publication.  In all the three cases, the title clearly indicates the scientific content continuity in the child paper.
Our specific contributions are summarized in the rest of this section.

\subsection{Reconciling obsolescence vs.\ entrenchment}

Our point of departure is the apparent contradiction between
obsolescence
\cite{Parolo2015734,ke2015defining,wang2013quantifying,Wang20094273,deSollaPrice510,ASI:ASI4630270505,leskovec2008microscopic}
and entrenchment
\cite{cho2005page,pandey2005shuffling,DBLP:journals/corr/VerstakASHILS14}.
We propose several measurements on evolving networks that constitute a \emph{temporal
bucket signature} summarizing the coexistence between entrenchment and obsolescence.
\textbf{Temporal bucket signature} denotes a stacked histogram of the
relative age of target papers cited in a source paper. Natural social
networks (e.g., various research communities) show diverse and
characteristic temporal bucket signatures.  Surprisingly, many
standard models of network evolution --- and even obsolescence ---
fail to fit the temporal signatures of real bibliometric data.  We
establish this with temporal bucket signatures and two associated
novel measures: \textbf{distance} and \textbf{turnover}. 
We also propose \textbf{age gap count histograms} to represent citation age distribution. Similar to temporal
bucket signature, standard models fail to fit age gap count histogram of real data as well. We establish this fitness using another novel metric termed as \textbf{divergence}. We define these in
Section~\ref{relay_main_results_signatures}.   As we shall see, simple
models with $O(1)$ parameters find it very challenging to pass all these
stringent tests for temporal fidelity.

\subsection{Insufficiency of intrinsic obsolescence}
Albert and Barabasi's remarkable scale-free model (preferential
attachment or \textbf{PA}) \cite{albert2002statistical} ``explained''
power law degrees, but failed to simulate many other natural
properties, such as bipartite communities.  The ``\textbf{copying}
model'' \cite{kumar2000random} gave a better power law fit and
explained bipartite communities.  Given that temporal signatures have not been studied before, it is not surprising that
these models fit real signatures poorly.  We demonstrate this in
Section \ref{sec:prior_models}.

Recent work \cite{leskovec2008microscopic,Wang20094273,wang2013quantifying}
has sought to remedy that classical network growth models do not capture aging.
Dorogovtsev \etal~\cite{dorogovtsev2000evolution} 
empirically showed that power law aging function better fits real citation networks. A similar study by Hajra \etal~\cite{hajra2005aging} reconfirms the previous claim. Additionally, they show the existance of two exponents and a possibility of a crossover from one to the other. Universally, the crossover value was roughly close to ten years after publication.  
Recently, Wang \etal~\cite{Wang20094273} modelled aging using an exponential decay function.
They propose that the probability of citing
paper $p$ at time $t$ is proportional to the product
$k_p(t)e^{-\lambda(t-b_p)}$, where $k_p(t)$ is the number of citations
$p$ has at time $t$, $b_p$ is its birth epoch, and $\lambda$ is a
global decay parameter.  We call this the \textbf{WYY} model, after
the authors.  To our surprise (Section~\ref{sec:prior_models}), \textbf{WYY} model improves only modestly upon PA or copying model at matching
age gap count histograms and temporal bucket signatures.

A more sophisticated model by Wang \etal~\cite{wang2013quantifying} involves three model parameters $\eta_p,
\mu_p, \sigma_p$ \emph{per paper}. In effect, this model is just a
reparameterization to achieve \emph{data collapse}~\cite{0305-4470-34-33-302} --- collapsing apparently
diverse citation trajectories into one standard function of age. We hypothesize that the reason
is that aging papers lose probability of getting cited, but none of the aging models \emph{use the graph structure} to predict where these citations
are likely to be redistributed.  This limitation also applies to
Hawkes processes~\cite{bacry2015mean,FarajtabarWGLZS2015}, which we
discuss in Section~\ref{sec:other_related_models}.

\subsection{Triad \underline{un}completion and relay-linking}

Triad completion (\emph{viz.}, if links $(u,v)$ and $(v,w)$ are present, consider adding $(u,w)$) has long been established~\cite{holme2002} as a cornerstone of link prediction.  The above observations led us to look for the \emph{reverse micro-dynamic pattern}: whether a popular older paper $p_0$, at a given time, starts losing citations in favor of a newer paper $p_1$ citing $p_0$.  Of course, we may only get to see the final decision to cite $p_1$ and not the process of ``dropping'' $p_0$.  Therefore, it is a delicate process to tease apart such ``relaying'' (from $p_0$ to $p_1$) effects from myriad other reasons for increase or decrease in popularity.  But we succeeded in designing high-precision filters that gathered strong circumstantial evidence that this effect is real (Section \ref{relay_si_evidence}).

This study led to a family of \textbf{relay-linking} models that are
the central contributions of this paper (Section~\ref{sec:relay_models}),
roughly speaking: to add a citation in a new paper, choose an existing
paper $p_0$, but if it is too old, walk back along a citation link to
$p_1$ and (optionally) repeat the process.  We call this hypothesized
process \emph{triad uncompletion} and the associated generative model
\emph{relay-linking}.

These  proposed relay-linking models or \emph{network influenced models} of aging mimic temporal signatures of real networks better than
state-of-the-art aging models.  In sharp contrast to existing work, we
avoid modeling aging as governed by network-exogenous rules or
distributions (whose complexity scales with the number of nodes).  Our
models have only two global parameters shared over all nodes. 

In Section~\ref{sec:dataset}, we describe a large-scale time-stamped bibliographic dataset. 
Section~\ref{sec:ObsolescenceEntrenchment} presents empirical evidences of co-existence of obsolescence and entrenchment,
leading to the development of the temporal bucket signatures described in Section~\ref{relay_main_results_signatures}. 
Section~\ref{sec:Classical} presents description of classical evolution models and our simulation framework.  In Section~\ref{relay_main_results_relay_models},
we present evidences of relay and propose several relay-linking models. We compare proposed relay-linking models in Section~\ref{relay_main_results_good_fit}.
Section~\ref{sec:TurnoverImpactFactor} presents an interesting application of the temporal bucket signatures.

\begin{figure*}[!tbh]
\begin{tabular}{@{}c@{}c@{}c@{}c@{}}
  \includegraphics[width=.25\hsize]{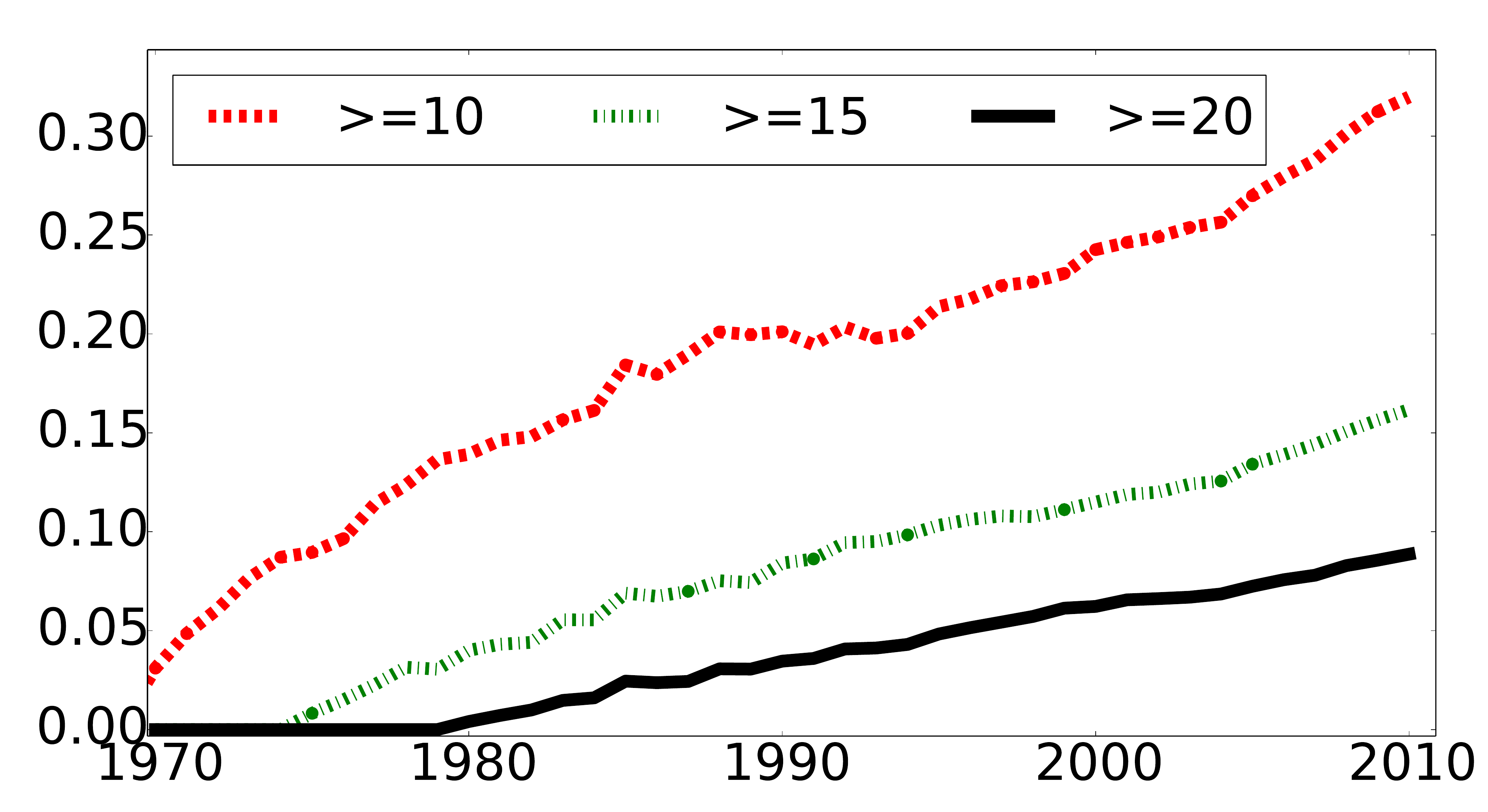} &
  \includegraphics[width=.25\hsize]{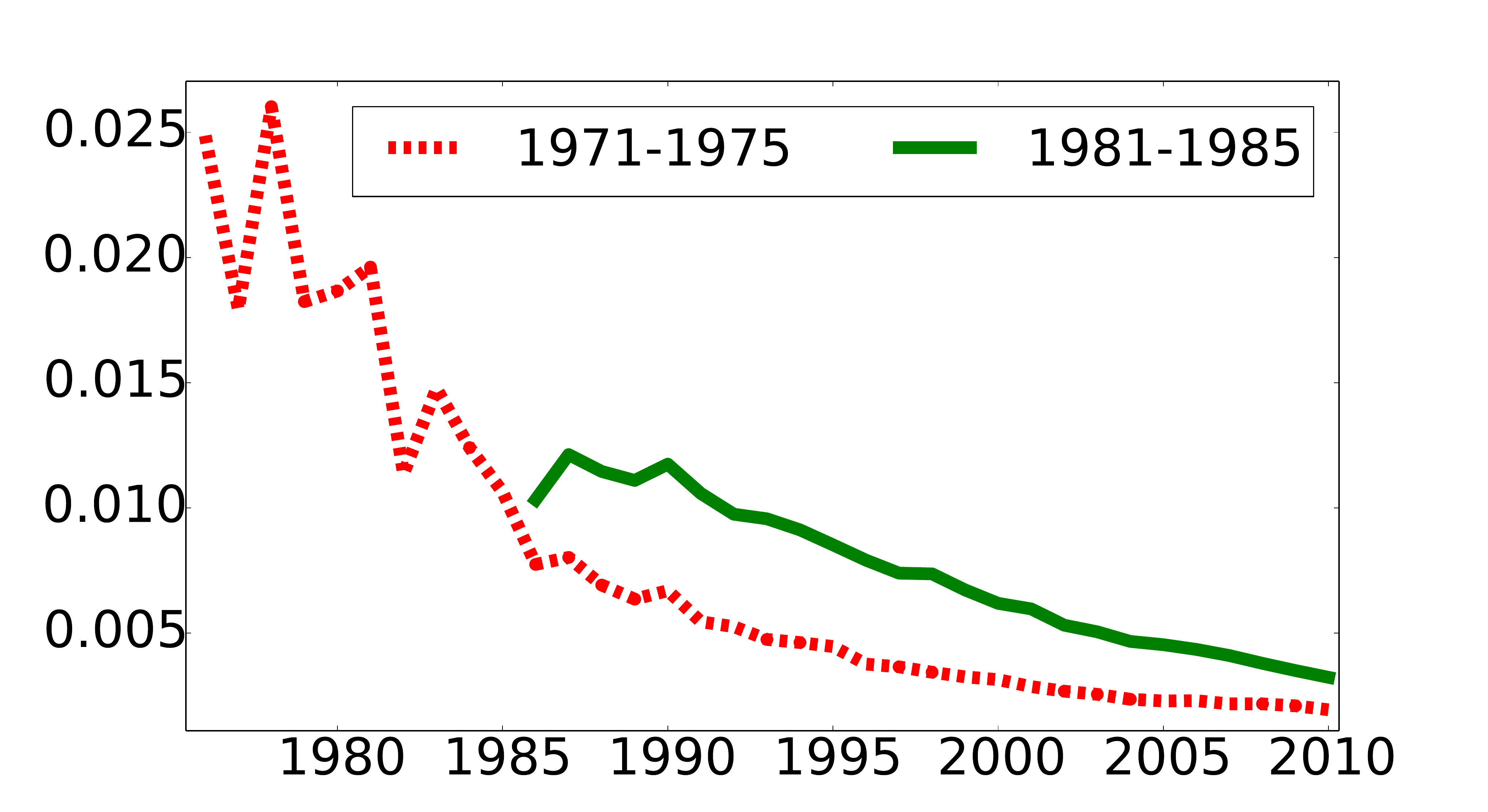} &
    \includegraphics[width=.25\hsize]{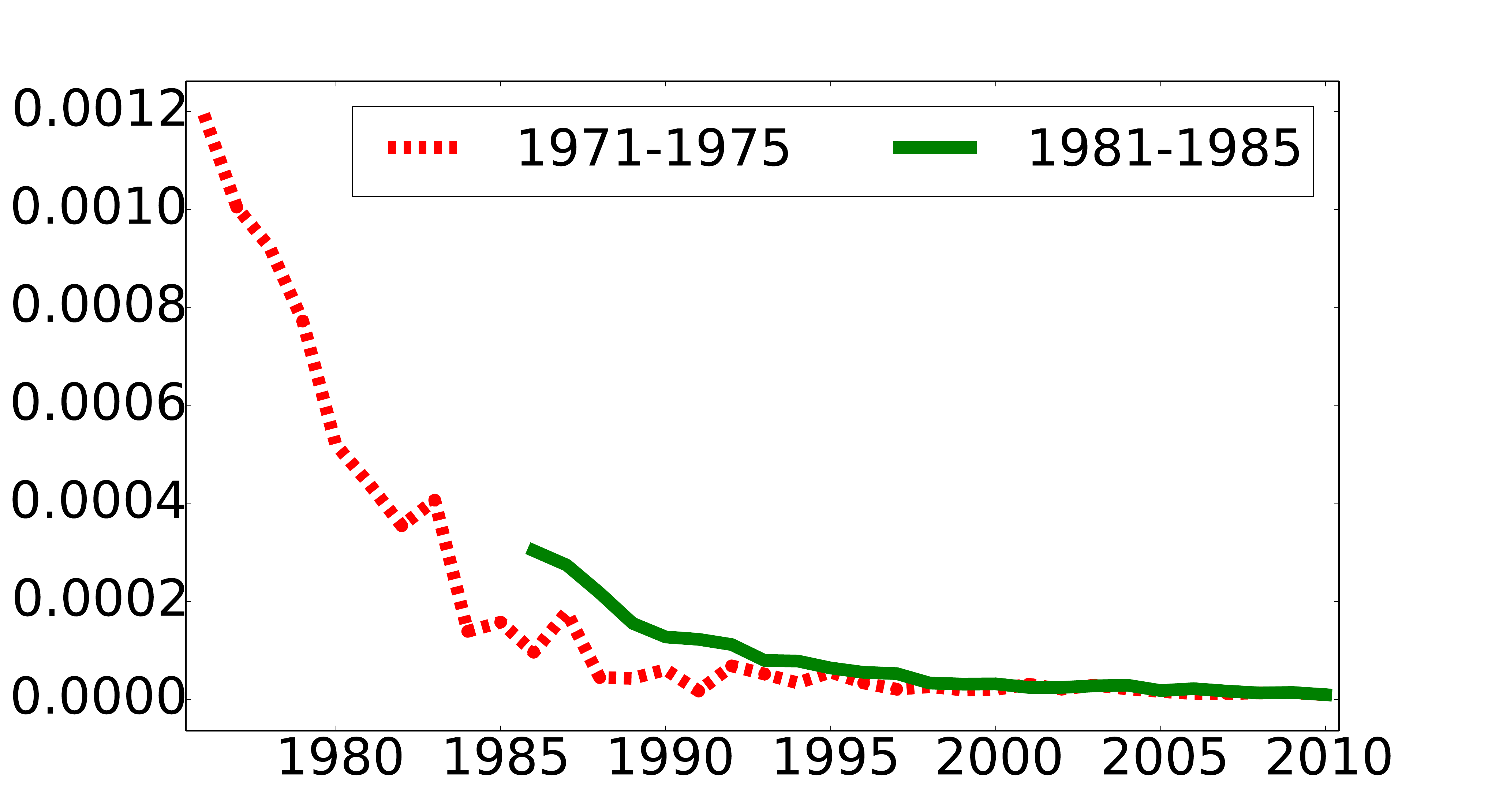} &
  \includegraphics[width=.25\hsize]{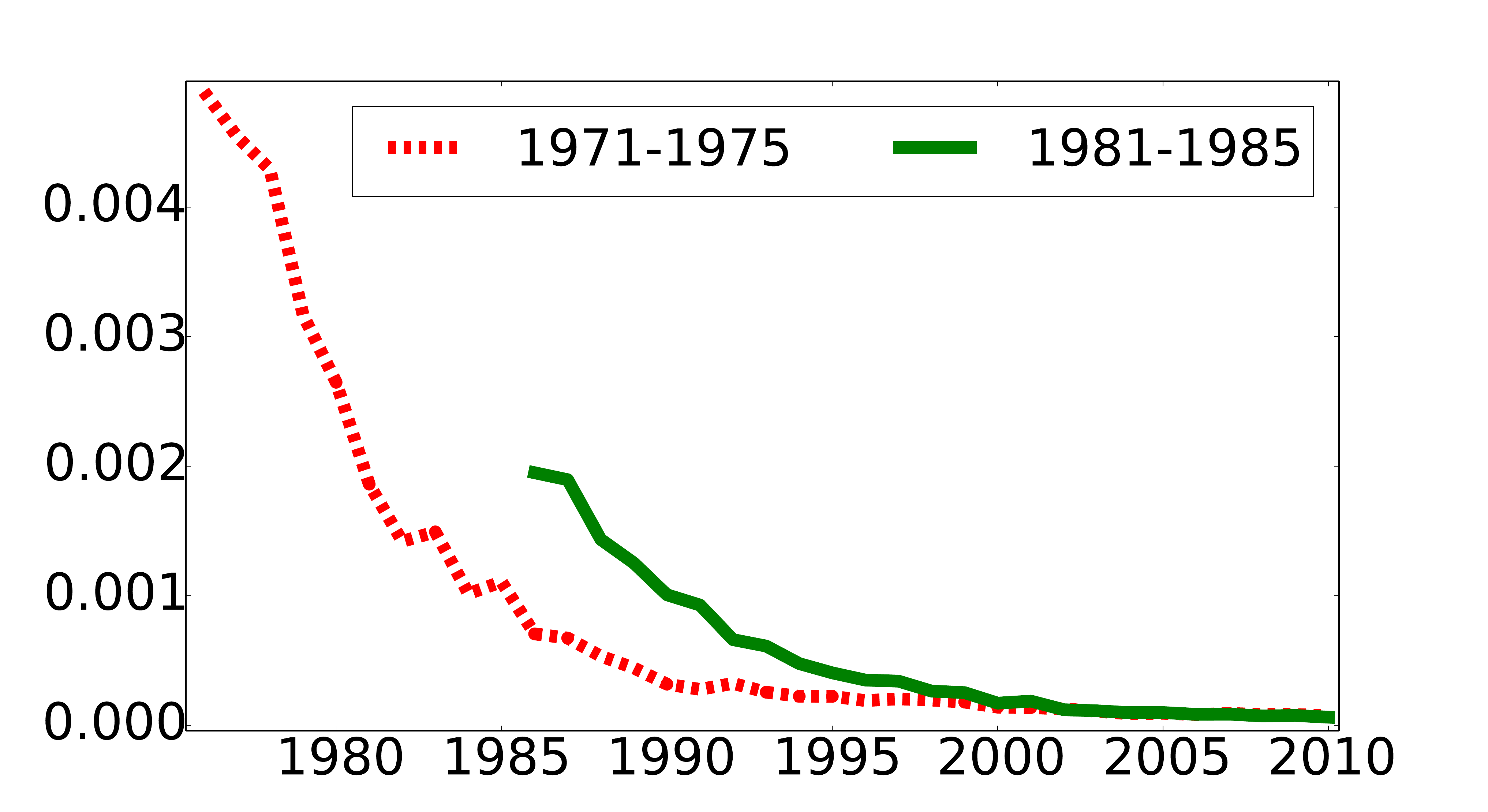} \\
  (a) & (b) & (c) & (d)\\
\end{tabular}
\caption{\textbf{(Best viewed in color) a.}~For a paper written in $y\in[1970,2010]$ (x-axis), we plot the fraction of papers it cites (y-axis) that are older than $y-t$ years, for $t = 10, 15, 20$ (red, green, black).  \textbf{b.}~We picked a fixed set $P$ of 100 most cited papers written in 1971--1975 (red) and 1981-1985 (green).  For papers written in years $y\in[1975,2010]$ (x-axis), we plot the fraction (y-axis) of citations made to papers in~$P$.  Unlike (a), this shows a steep decrease.  \textbf{c.}~Replacing popular papers $P$ with a random set $R$ of papers written in 1971--1975 (red) and 1981--1985 (green) reduces the \emph{absolute} y-axis but not the \emph{relative} decay.  \textbf{d.}~Enlarging $R$ to 500 random papers also has no effect on the relative rate of decay.}  \label{fig:citations}
\end{figure*}

\section{Dataset}
\label{sec:dataset}
Investigating the questions raised in this work requires rich trajectories of
time-stamped network snapshots.  However, such intricately detailed
datasets are rare, even while there are an increasing number of new
repositories being built and updated
regularly\footnote{\protect\url{http://snap.stanford.edu/} is a
  prominent example.}.  Fortunately, Microsoft Academic
Search\footnote{\protect\url{http://academic.research.microsoft.com}}
(MAS) provides an ideal platform for our study.  MAS data includes
paper titles, reconciled paper IDs, year of publication, publication
venue, references, citation contexts, related field(s), abstract and
keywords, author(s) and their affiliations~\cite{Chakraborty:2014}.
We have filtered papers from full dataset (Table \ref{tab:dataset_I}).
The filtered dataset consists of papers published between 1961--2010 and have at least one outlink or one inlink
(to filter isolated nodes or missing data). We call this filtered dataset as the Ground Truth dataset (GT). For each simulation initialization, 
we create a warmup dataset from GT having papers published between 1961--1970. Detailed description and the role of warmup data in the simulation framework can be found in Section~\ref{sec:prior_models}.

\begin{table}[ht]
\centering \normalsize
  \caption{General statistics about the full Computer Science dataset
    from Microsoft Academic Search. Filtered and warmup dataset are subsets of full dataset.} \label{tab:dataset_I}
  \begin{tabular}{|c|c|c|c|}
    \hline
   &Full&Filtered&Warmup \\ \hline
  Year range&1859--2012 & 1961--2010 & 1961--1970 \\ \hline
  Number of papers &2,281,307 &1,702,471& 9,568 \\ \hline
  Number of citations & 27,527,432 & 15,791,272 & 7,312\\ \hline
 \end{tabular}
\end{table}

To ensure that our proposed temporal signatures are generally applicable, we also experimented with papers from the biomedical domain. In this study, we use biomedical dataset that consists of 801,252 research articles\footnote{\protect\url{http://www.ncbi.nlm.nih.gov/pmc/tools/ftp}} published between 1996-2014.  All our evaluations are based on extensive experiments with the Computer Science domain dataset\footnote{We have comparable evaluation on biomedical papers which we omit due to space constraints.}.

\section{Entrenchment and obsolescence}
\label{sec:ObsolescenceEntrenchment}
Preferential attachment models without aging
\cite{jeong2003measuring,kumar2000random} predict that older papers
get more entrenched and their rate of citation acquisition can only go
up.  Verstak \etal\ \cite{DBLP:journals/corr/VerstakASHILS14} provide
support that \emph{as a cohort} older papers are thriving: more
recently written papers have a larger fraction of outbound citations
targeting papers that are older by a fixed number of years.  However,
there is plenty of evidence
\cite{Chakraborty:2015:CSC:2817191.2701412,wang2013quantifying,Wang20094273}
that aging counteracts entrenchment.  This apparent contradiction is
readily resolved by realizing that the number of papers older by a
fixed number of years is growing rapidly.  But the real value of the
study (Sections~\ref{sec:frac-cite-old} and
\ref{sec:frac-cite-bucket-10}) is that it leads us to the definition
of new signatures of evolving networks
(Section~\ref{relay_main_results_signatures}).

\subsection{Fraction of citations to `old' papers}
\label{sec:frac-cite-old}

Suppose that papers in our corpus, published in year $y$, make $C_y$ citations in all to older papers.  Of these, say $C_t$ citations go to papers that were published before year $y-t$, for $t = 10,15,20$.  Figure~\ref{fig:citations}(a) plots the quantity ${C_t}/{C_y}$  against~$y$, similar to the setup of Verstak et al.~\cite{DBLP:journals/corr/VerstakASHILS14}.  The plot is consistent with their claim: the fraction of citations to older papers is indeed increasing over the years $y$ for all values of $t$.

However, Figure~\ref{fig:citations}(b) paints a different picture.
For each year range 1971--1975 and 1981--1985, we choose 100 most cited (through 2010) papers $P$.  Then, for other papers written in year $y \in [1975,2010]$, we plotted the fraction of citations out of those papers that go to~$P$.  Clearly, this fraction  decreases over time.  In place of popular papers, how do \emph{random} papers fare?  Figures~\ref{fig:citations}(c,d) show that the relative shape of decay remains stable when random paper sets of sizes 100 and 500 are picked as the targets.

\subsection{Fraction of citations to papers in 10-year age buckets}
\label{sec:frac-cite-bucket-10}

\begin{figure}[ht]
  \centering
  \includegraphics
      [width=.85\hsize]{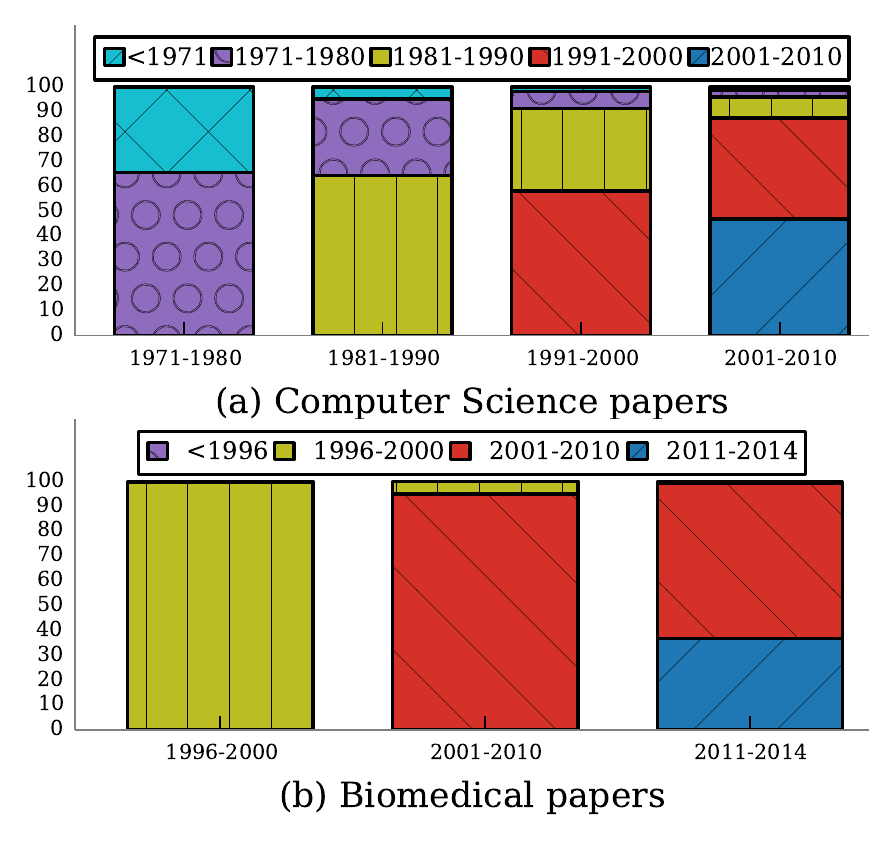}
      \caption{\textbf{a.}~Citation distribution across 10-year buckets for computer science dataset.  Each
        vertical bar represents a decade of papers.  Within each bar,
        colored/textured segments represent the fraction of citations
        going to preceding decades.  The bottommost segment is to the
        same decade, the second from bottom to the previous decade,
        etc.  On one hand, the volume of citations to the current
        decade (bottommost segment) is shrinking to accommodate ``old
        classics'' (entrenchment).  On the other hand, any given
        color/texture shrinks dramatically over decades (most papers
        fade away). \textbf{b.}~Citation distribution for biomedical dataset. Papers written
	in 1996--2000 became obsolete much more rapidly. }
\label{fig:buckets_all}
\end{figure}

Figures~\ref{fig:citations} suggests a natural and compact way to
summarize citation statistics organized by age.  We group papers into
buckets.  Each bucket includes papers published in one decade\footnote{Any
suitable bucket duration can be used. We experiment with several bucket sizes, majority of them produced similar results.}.  Then, for each bucket, we plot
as a stacked bar-chart, the fraction of citations going to that same
bucket as well as all previous buckets.
Figure~\ref{fig:buckets_all}a shows the result.  We note the following:

\begin{itemize}
\item The fraction of citations from a bucket to itself (shown as the bottom purple, yellow, red and blue bars in successive columns)
decreases over time, and those to all older buckets increases over time.  This is consistent with Verstak \etal{}
\item However, if we consider papers in a bucket as targets, the citations they receive decreases over the years.
For instance, papers written in 1971--1980 (purple bars over successive columns) received 70.5\% of the citations in
that decade (purple) but this number reduces to $29.2,6.4,2.8$\% in successive decades.  Similar decay is seen for the following buckets (yellow, red) as well.
\end{itemize}

We see similar effects in Figure~\ref{fig:buckets_all}(b), except that papers written
in 1996--2000 became obsolete much more rapidly (yellow bar) compared
to papers written in 2001--2010, so there is less stationarity of the
obsolescence process in the biomedical domain compared to computer
science. Thus, such bar charts \emph{simultaneously} validate Verstak \etal~\cite{DBLP:journals/corr/VerstakASHILS14} and also show aging of paper cohorts, and are a succinct signature of the balance between entrenchment and obsolescence.

\begin{figure*}[!tbh]
  \centering
  \adjustbox{valign=c}{\includegraphics[width=.7\hsize]{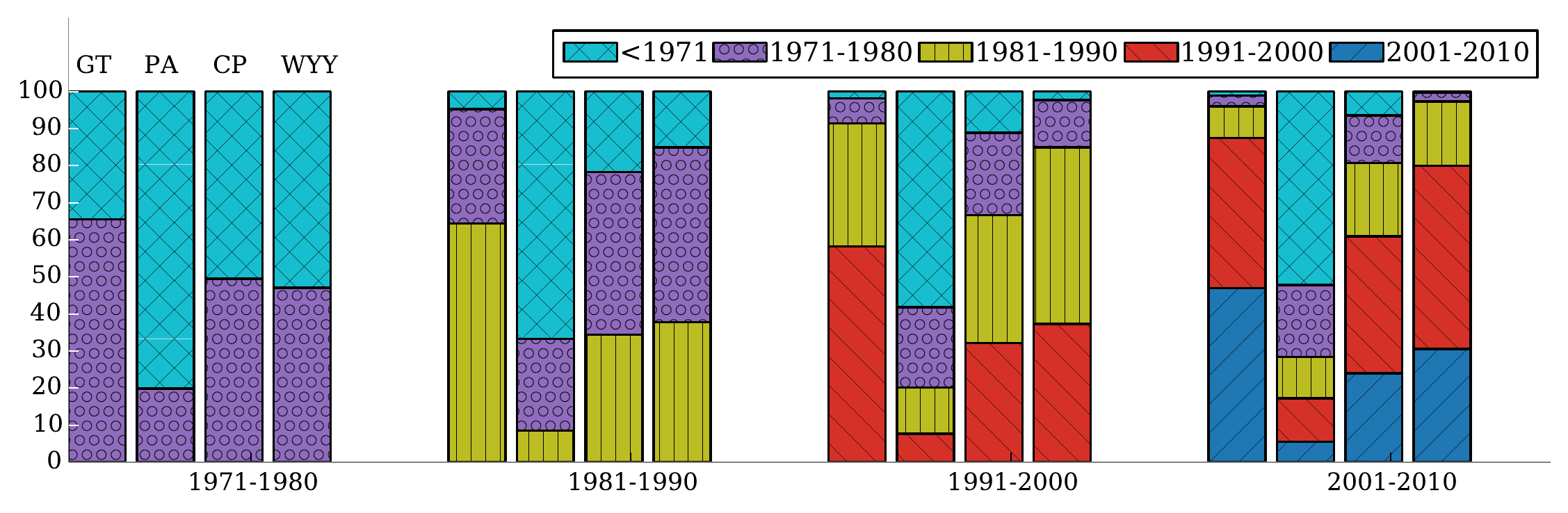}}
  \adjustbox{valign=c}{
  \begin{tabular}{l|r|r|r}
  Simulator & \rotatebox{90}{Distance} &
  \rotatebox{90}{Turnover} & \rotatebox{90}{Divergence} \\ \hline
  GT &--&2.70&--\\
  PA & 4.98 & 0.97 & 0.77 \\
  CP ($p_\text{copy}=.5$) & 1.97 & 1.53 & 0.18 \\
  WYY ($\lambda=.11$) & 1.67 & 2.59 & 0.13
  \end{tabular}}
  \caption{Temporal bucket signatures comparing ground truth (GT),
    preferential attachment (PA), copying (CP) and WYY, the model proposed by Wang 
    \etal~\cite{Wang20094273}.  Each
    bucket represents a decade.  Ground truth turnover is 2.70.  For
    others, distance, turnover and divergence values are shown in the accompanying table.
    Clearly, only WYY has even a remote similarity to ground truth.}
  \label{fig:GT_PA_CP_WYY}
\end{figure*}

\section{New signatures of evolving networks}
\label{relay_main_results_signatures}

We start with some basic notation. Time $t$ proceeds in
discrete steps (for publications, often measured in years). 
Sometimes we will bucket time into ranges like
decades. We study an evolving graph $G_t$, which comprises the node
set $V_t$ and edge set $E_t$.  Nodes are denoted by $u, v$, etc.  Edges
(i.e., citations) once added, are never removed.  Also, in our
bibliometric setting, edges emanating from a node $v$ all ``appear''
when node $v$ itself appears, at birth time $t_v$, but this assumption
can be relaxed. We shall use GT as the shorthand for ground-truth data (see Section~\ref{sec:dataset}). 

We introduce several natural ways to observe dynamic networks to better
understand the interplay between entrenchment and obsolescence.

\subsection{Age gap count histogram}
\label{sec:Reconcile:AgeGapCount}

When new paper $u$, born at time $t_u$, cites an older paper $v$, born at
$t_v$, that citation link spans an \emph{age gap} of $t_u - t_v \ge
0$.  (Depending on the granularity of measuring time, $t_u=t_v$ may or
may not be possible.)  In case of dynamic documents where $u$ can add
citations (dropping citations is rare), we can take $t_u$ to be the
citation creation time, rather than the birth time of~$u$.  In
citation data, gap $g$ is usually expressed in whole years.  For any
value of $g$,
\begin{align}
  \sum_{(u,v)\in E} 
  \begin{cases}
    1, & \text{if $t_u - t_v = g$, and} \\
    0, & \text{otherwise}
  \end{cases}  \label{eq:AgeGapCount}
\end{align}
is the number of links that span an age gap of~$g$.  As we shall see later, age gap count histograms
reveal some salient dynamics of graph evolution.

\subsubsection{Divergence}
Suppose we observe age gap histograms $H$ from real data.  
Each simulated model gives age gap histograms  $\tilde{H}$.  
We assess divergence between two histograms ($\tilde{H}$ and $H$) by measuring
Kullback-Leibler divergence. More precisely,
\begin{align}
  \text{divergence}(H||\tilde{H}) = \sum_{g\in H} H(g)\log{\frac{H(g)}{\tilde{H}(g)}}
\end{align}
A simulated model is closer to real data, if $\text{divergence} \rightarrow 0$.

\subsection{Temporal bucket signature}
\label{sec:fitting_metric}

Suppose we collect birth times into buckets of temporal width $T$ (e.g., $T$ may be 10 years).  Suppose our corpus of papers $P$ is thus
partitioned into $P_1, P_2, \ldots, P_N$, based on their publication
date.  We pad this with sentinel bucket $P_0$ for all papers before
$P_1$.  Each source paper $p_s \in P_j$ may cite target papers $p_t
\in P_i$, where $i \le j$.  Let the total number of citations from
papers in $P_j$ to papers in $P_i$ be $C(i,j)$ (row=cited,
column=citing).  Let column sums $C(j) = \sum_i C(i,j)$ be the total
number of outbound citations from papers in $P_j$.  Let $F(i,j) =
C(i,j) / C(j)$ be the fraction of outbound links from papers in $P_j$
that target papers in $P_i$.  The temporal bucket signature is defined as
the matrix $F(i,j): i \le j$, i.e.,
\begin{align}
  F &=  \begin{bmatrix}
    F(0,1) & F(0,2) & \cdots & F(0,N) \\
    F(1,1) & F(1,2) & \cdots & F(1,N) \\
    0      & F(2,2) & \cdots & F(2,N) \\
    \vdots & 0      & \ddots & \vdots \\
    0      & 0      & 0      & F(N,N)
  \end{bmatrix},
\end{align}
where each column adds up to 1.  We propose two intuitive scalar
summaries of temporal bucket signatures.

\subsubsection{Distance}

Suppose we observe $F$ from real data.  We also fit a model which,
upon simulation, gives bucket signature $\tilde{F}$.  We propose to
assess how closely $\tilde{F}$ approximates $F$ by measuring the
average row-wise L1 distance between their corresponding columns.  More
precisely,
\begin{align}
  \text{distance}(F, \tilde{F}) &=
   \sum_{j = 1}^N \left[
    \sum_{i=0}^j |F(i,j) - \tilde{F}(i,j) |
    \right].
\end{align}
The higher the distance value, lower will be the closeness of approximation, and vice versa. 
Note that there is no assumption
of stationarity in this definition.  Communities can be in
volatile and transient stages of obsolescence while replacement rates
in other communities can be stable.

\subsubsection{Turnover}
\label{sec:Reconcile:Turnover}

Another quantity of interest summarizing $F$ or $\tilde{F}$ is a
notion of \emph{decay} of the height of a segment of a given color
from one column to the next, in the sequence $F(i,i), F(i,i+1),
F(i,i+2), \ldots$ Specifically, the ratio $F(i,j)/F(i,j+1)$ (which is
usually more than 1) represents how sharply citations to papers in
$P_i$ decreases from year $j$ to year $j+1$.  Because we are
interested in a ratio, we aggregate these via a geometric mean:
\begin{align}
  \text{turnover}(F) &=
  \left[
    \prod_{j=1}^{N-1} \prod_{i=0}^{j} \frac{F(i,j)}{F(i,j+1)}
    \right]^{\frac{2}{(N+2)(N-1)}}
\end{align}
A high value of \emph{turnover} indicates more rapid obsolescence.
Turnover can be measured on both $F$ and~$\tilde{F}$.  In the later sections,
we will relate the quantities we have defined with other established properties of real networks.

\subsection{Optimization}
\label{sec:optimisation}
We assume that the temporal bucket signature for GT is $F$ and the age gap histogram is $H$. Similarly, for each simulated model, we denote $\tilde{F}$ and $\tilde{H}$ as temporal bucket signature and age gap histogram respectively. Note that, $\tilde{F}$ and $\tilde{H}$ 
are dependent on two model parameters $\lambda$ and $\Theta$ (see Figure~\ref{fig:relay-link-template}). We use $d(\cdot)$, $t(\cdot)$ and $f(\cdot)$ as shorthand for $\text{distance}(\cdot)$, $\text{turnover}(\cdot)$ and $\text{divergence}(\cdot)$ respectively. To obtain optimal set of parameters for each model, we need to solve the following optimization problem:
\begin{equation}
\label{eqn_optimize}
\begin{aligned}
& \underset{\lambda,\theta}{\text{minimize}}
& & \emph{d}(F, \tilde{F})*\left(|t(\tilde{F}) - t(F)|\right)*f(H||\tilde{H}) \\
\end{aligned}
\end{equation}
Here, $|t(\tilde{F}) - t(F)|$ represents absolute difference between GT's turnover 
(e.g., 2.70 for one of our data sets), and relay-link model's turnover. Other combinations such as weighted sums can be considered, but product has the advantage that we do not need to manually balance typical magnitudes of the parts.
To our knowledge the above problem does not admit a tractable continuous optimization procedure.
Therefore, we perform grid search and choose values for model parameters for each proposed model.

\section{Classical evolution models and simulation results}
\label{sec:Classical}

The first generation of idealized network growth models
\cite{albert2002statistical,pennock2002winners} generally focused on a
``rich gets richer'' (preferential attachment or PA) phenomenon
without any notion of aging.  This was followed by the vertex
copying model \cite{kumar2000random}. There has been more recent work
\cite{dorogovtsev2000evolution,hajra2005aging,wang2013quantifying,Wang20094273,zhu2003effect} on modeling age
within the PA framework.  We will review and evaluate some of these in
Section~\ref{sec:prior_models}.

\subsection{Classical Models}
\subsubsection{Standard preferential attachment (PA)}
\label{sec:pref-attach}

In Albert \etal's classical PA model \cite{albert2002statistical,jeong2003measuring}, at time $t$, a new paper would cite an 
old paper $p$, which currently has degree $k_p(t)$, with probability $\Pi(p,t)$ that is proportional to $k_p(t)$:

\begin{equation}
\Pi(p,t) \propto k_p(t)  \label{eq:pa}
\end{equation}

In their idealized model, one new paper was added at every time step, but this is easily extended to mimic and match the growing observed rate 
of arrival of new papers.  Moreover, the number of outbound citations from each new paper can also be sampled to match real data.

If paper $p$ arrives at time $t_p$, it is not hard to obtain a mean-field approximation to the degree of $p$ at time $t \ge t_p$:

\begin{align}
  \tilde{k}_p(t) \propto \sqrt{ t / t_p }.
\end{align}

This expression suggests that age is a monotone asset, never a
liability, for any paper. 

\subsubsection{Copying model (CP)}
\label{sec:vertex-copy}

The copying model \cite{kumar2000random} is characterized by a
network that grows from a small initial graph and, at each time step,
adds a new node (paper) $p_n$ with $k$ edges (citations) emanating
from it.  Let $p_r$ be a ``reference'' paper chosen uniformly at
random from pre-existing papers.  With a fixed probability (the only
parameter of the model), each citation from $p_n$ is assigned to the
destination of a citation made by $p_r$, i.e., $p_n$ ``copies''
$p_r$'s citations.  Neither PA nor copying has a notion of aging.
  
\subsubsection{Ageing model (WYY)}
Wang, Yu and Yu~\cite{Wang20094273} proposed modeling age within the PA framework. The probability of citing at time $t$ a paper $p$ that
was born at time $b_p$, while proportional to its current degree as in
PA, \emph{decreases exponentially} with its age:
\begin{align}
  \Pi(p,t) &\propto k_p(t)\,\exp\Bigl(-\lambda (t - b_p) \Bigr),
  \label{eq:ia1}
\end{align}
where $\lambda>0$ is the single global parameter controlling the
attention decay rate, estimated from some ``warmup'' data. Similar
models are motivated by the measurements by Leskovec
\etal~\cite{leskovec2008microscopic}. Note, in order to avoid huge computational overhead associated with updating  probability values for each  new entry, we approximate by only updating the attachment probability value once in each year. For the first 20 years, the approximate version is (a) extremely close to the original version (less than .05 L1 distance) and (b) slightly closer to the GT than the original version thus giving this baseline a small additional advantage.

\subsection{Simulation protocol and results}
\label{sec:prior_models}
We simulate the models described above for 40 years (1971--2010) and compare the results with GT ($turnover=\textbf{2.70}$).  
\textit{Warmup data} is the subset of GT generated between 1961--1970 (detailed statistics is present in Section~\ref{sec:dataset}). 
Warmup data consists of papers published between 1961--1970 along with the citation links formed between them. We initiate each 
simulation model from warmup data. The warmup data can be called as the ``train data''. Starting from the year 1971,
for each subsequent year we introduce as many papers in the system as the publication count of that year estimated from GT. Each incoming paper
is accompanied by nine outlinks (average number references estimated from GT). This data, generated through our simulation models
between 1971--2010, can be called as ``test data''.
We simulate CP with copying probability = 0.5 (after grid search on all possible probability values) since the product of the three observables,
i.e., distance, turnover, and divergence (a function similar to~\eqref{eqn_optimize}) is the least at this value of the probability. Similarly,
for WYY, we obtain through grid search $\lambda = 0.11$ that results in the lowest product of the three observables.

Results are shown in Figure~\ref{fig:GT_PA_CP_WYY}.  PA fits
observed temporal bucket profiles very poorly. The distance score is very large (\textbf{4.98}). Neither PA nor copying has a notion of aging.
Therefore, it is not surprising that CP also does not fit observed
temporal bucket signatures well.  The distance score is 
\textbf{1.97}. WYY performed best at $\lambda = 0.11$ with distance = \textbf{1.67}. 
As for turnover, WYY's turnover (\textbf{2.59}) is closest to that of GT (\textbf{2.70}).

\subsection{Other related models}
\label{sec:other_related_models}

\subsubsection{Forest Fire}
Relay-linking has some superficial similarity to the forest fire model
\cite{leskovec2005densification} and earlier work on random walk and
recursive search based attachment processes
\cite{Vazquez2001recursive}.  But among many critical difference is
the involvement of time and node ages.  In forest fire terminology,
the relative birth times of candidate source and target nodes strongly
influence whether we prefer to `burn' forward or backward edges. To
our knowledge, there is no similar temporally modulated version of
forest fire model that has demonstrated fidelity to bucket signatures, or age gap count histograms. 

\subsubsection{Point processes}
It is attractive to think of citations as events ``arriving at a
node/paper'' according to some temporal point
process\footnote{\protect\url{https://en.wikipedia.org/wiki/Point_process}}.
Focusing on one node, if $\mathcal{H}(t)$ is the history of event
arrivals up to time $t$, then the \emph{conditional intensity
  function} is defined as
\begin{align*}
  \gamma(t) dt &:= \Pr(\text{event in $[t, t+dt)$} | \mathcal{H}(t)).
\end{align*}
Specifically, if $\mathcal{H}_v(t)$ comprises the points of time
${t_{vi} < t}$ of past arrivals at node $v$, then the Hawkes process
\cite{aalen2008survival} defines
\begin{align*}
  \gamma_v(t) &= a_v + b_v \sum_{t_{vi} < t} \exp(-|t - t_{vi}|).
\end{align*}
and provides two major benefits:
\begin{enumerate*}
\item the exponential decay term elegantly captures temporal
  burstiness, and
\item given $\{t_{vi}\}$, parameters $a_v, b_v$ can be estimated
  efficiently \cite{bacry2015mean,FarajtabarWGLZS2015}.
\end{enumerate*}
While Hawkes process is most suited for repeated similar events (such
as messages or tweets between two people), citation happens only once
between two papers.  Work on coupling edge message events to network
evolution itself is rare, with notable exceptions
\cite{FarajtabarWGLZS2015}.  In our case, citation arrivals at
different papers are not independent events, but coupled to global
population growth rates as well as network constraints (e.g.,
out-degree distribution).  Given those constraints, Hawkes process
provides no obvious benefits to inference or simulation.  Moreover,
citations are often observed in (annual) batches, but Hawkes process
finds simultaneous arrivals impossible.  We can model arrival times as
hidden and observe them in batches, but that involves a more complex
EM procedure \cite{liu2015efficient} to marginalize over arrivals.
Even if these hurdles can be overcome, we have to estimate or sample
$a_v, b_v$ for every node, just like WSB \cite{wang2013quantifying},
which results in too many parameters.  Moreover, there is still no
direct connection between declining citations and whether the network
guides the diverted citations to specific targets, which is the
specific goal of relay-linking models.

\section{Proposed relay-linking models}
\label{relay_main_results_relay_models}

\subsection{Evidence of citation stealing}
\label{relay_si_evidence}

The central hypothesis behind the relay linking model is as follows:

\begin{mdframed}[backgroundcolor=gray!10,font=\bfseries] 
At a given point in time, an old popular paper $p_0$ begins to lose
citations in favor of a relatively young paper $p_1$ that cites $p_0$.
\end{mdframed}
There are a variety of intuitive reasons why relay-linking or relay-citing can happen:
\begin{itemize}
\item $p_1$ is a journal version of a conference paper $p_0$,
\item $p_1$ refutes or improves upon $p_0$, or
\item $p_1$ reuses data or a procedure in $p_0$, and so on.
\end{itemize}

\begin{table*}[!htb]
\centering
\caption{Circumstantial evidence of relay-link:
  $R_W$ papers acquire more citations than $R_L$ papers. Here, $r$ is in $R_W$ or $R_L$ . Higher
proportion of papers belonging to $R_L$ have zero citation count than $R_W$. Bold face text represents that the mean of cumulative citation
count of $R_W$ at base year $T$ is larger than the mean of $R_L$. Also, $R_W$ papers show higher increasing trend than $R_L$ papers.}
\label{tab:relay-evidence}
\begin{tabular}{|c|r|c|c|c|c|c|c|c|c|}
\hline
& \parbox[t]{1.3cm}{\raggedright Popularity of cited papers} & \parbox[t]{1cm}{\#Papers} &\parbox[t]{1.2cm}{\raggedright \#Papers with $>0$ citations} & \parbox[t]{1.2cm}{\raggedright \%Papers with $>0$ citations} & \parbox[t]{1.2cm}{\raggedright Avg \#citations to $r$} & \parbox[t]{1.8cm}{\raggedright \#Recent papers with increasing trend} & \parbox[t]{2cm}{\raggedright \%Recent papers with increasing trend} & \parbox[t]{2.1cm}{\raggedright Cited neighbors with decreasing trend (\%)} & \parbox[t]{1.5cm}{\raggedright Avg. decrease in median values} \\\hline
$R_W$& $\ge 70$ & 76082 & 60205 & \textbf{79.13}& 19.77& 31749& \textbf{41.72} & 48.06& 5.69\\\hline
$R_L$& $\le 10$ & 16257 & 2017  & \textbf{12.40} & 0.31& 736& \textbf{4.52}& 41.39 & 0.41\\\hline
\end{tabular}
\end{table*}

\begin{table*}
\centering
\caption{Circumstantial evidence of relay-link: Papers that cite fading papers gather citations at an accelerated pace. Bold face text
represents that the rate at which the citations are gained by the set of $R'$ papers is higher compared to the set of $R_W \setminus R'$ papers.}
\label{tab:relay-evidenceII}
\begin{tabular}{|l|l|l|l|l|l|l|l|l|}
\hline
\multicolumn{3}{|c|}{$R_W$}&\multicolumn{3}{|c|}{$R'$}&\multicolumn{3}{|c|}{$R_W \setminus R'$}\\\hline
\parbox[t]{1cm}{\raggedright \#papers in $P_P$} & \parbox[t]{1cm}{\raggedright \#papers in F} &\parbox[t]{1cm}{\raggedright Avg. drop} & \parbox[t]{1.7cm}{\raggedright Avg. citation count at T} & \parbox[t]{2.3cm}{\raggedright Avg. citation gain in $[T, T+\delta T]$} & \parbox[t]{1.5cm}{Per-year citation gain} & \parbox[t]{1.7cm}{\raggedright Avg. citation count at $T$} & \parbox[t]{2.1cm}{\raggedright Avg. citation gain $[T, T+\delta T]$} & \parbox[t]{1.8cm}{\raggedright Year-wise citation gain}\\\hline
21621& 4962 & 36.41 & 23.48 & 13.92&\textbf{2.48}& 11.02& 11.89& \textbf{2.05}\\\hline
\end{tabular}
\end{table*}

Unlike standard preferential attachment (PA), evidence for relay-linking can only be circumstantial and in the aggregate, because the
decision of $p_2$ to select, but then \emph{not} cite $p_0$, is never recorded in any form; we get to know only of the recorded citation
to $p_1$.  Here we produce such circumstantial evidence, in two parts.

Fix a base time $T$ (2005 in our experiments).
Define \textbf{popular} papers $P_P$ as those that have at least 70 cumulative citations as of $T$.  Define \textbf{obscure} papers $P_O$ as
those that have at most ten cumulative citations as of $T$.  Let \textbf{recent winner} papers $R_W$ be those that make at least ten 
citations\footnote{To eliminate noise in extracting citations.} and at least  50\% are to papers in $P_P$.  Let \textbf{recent loser} 
papers $R_L$ be those that make at least ten citations, and all are to papers in $P_O$.

\paragraph{Do $R_W$ papers gain citations faster than $R_L$?}
We now measure the cumulative citations to each paper in $R_W$ and $R_L$ as of time $T+\delta T$ (say after five years), and can apply a standard test of the hypothesis that the mean of $R_W$ is larger than the mean of $R_L$
(see Table~\ref{tab:relay-evidence}). 

\paragraph{Are $R_W$ papers stealing citations from $P_P$ papers?}
Now we focus on a subset of $P_P$: those whose rate of acquiring citations see a sharp ($>50\%$) drop from $[T-\delta T, T]$ to $[T, T+\delta T]$.  Let this be \textbf{fading} papers $F \subset P_P$.  Consider papers  $R' \subset R_W$ that cite papers in $F$, and their rate of acquiring citations in $[T, T + \delta T]$.  We investigate if this population has a significantly larger mean than a base population. Here the base population are set to the papers $R_W \setminus R'$. In Table~\ref{tab:relay-evidenceII} we observe that indeed the rate at which the citations are gained by the set of $R'$ papers is higher compared to the set of $R_W \setminus R'$ papers.

\subsection{Model descriptions and results}
\label{sec:relay_models}
Inspired by the above experiments, we propose in Fig.~\ref{fig:relay-link-template}, a generic template for all
our relay-link models. $t_u$ is the birth time of $u$.  The flexible policies/ parameters are $R, \lambda, \Theta, D$.  $R$ is either 1 (one-shot relay) or $\infty$ (iterated relay). $D$ is either uniform, or as in PA, but restricted to $I(u,t)$. The $\lambda$ parameter governs the time to initiate relaying while  the $\Theta$ parameter governs the extent of relaying. Higher value of $\lambda$ leads to relaying of citations from source paper soon after its publication and vice-versa. Similarly, $\Theta$ controls intensity of relaying; higher values of $\Theta$ lead to higher intensity of relaying. Note that, standard PA can be achieved by keeping $\lambda =0$. We will explore alternatives for a few design
choices, and that will lead us to a few variations on the basic theme.

\subsubsection{Random relay-cite (RRC)}
\label{sec:Random_Delinking_Model}

  Our first model is obtained by setting $R=1$ and $D$ as the uniform distribution over $I(u,t)$.  In words, we first pick a $p_0$ to cite, then we toss a coin with head probability = $exp(-\lambda T)$, where T is the current age of the paper $p_0$. If the coin turns up tail, then again, we toss a coin with head probability $\Theta$.  With coin turning up as head, we sample a paper $v$ that links to $p_0$ uniformly at random, and then cite $v$ instead of $p_0$.  Effectively, $p_0$ \emph{relays} the citation to $v$.  This version of the model thus has two parameters $\lambda$ and $\Theta$.

\begin{figure*}[!htb]
  \centering \hspace{-0.6cm}
   \adjustbox{valign=c}{\includegraphics[width=.7\hsize]{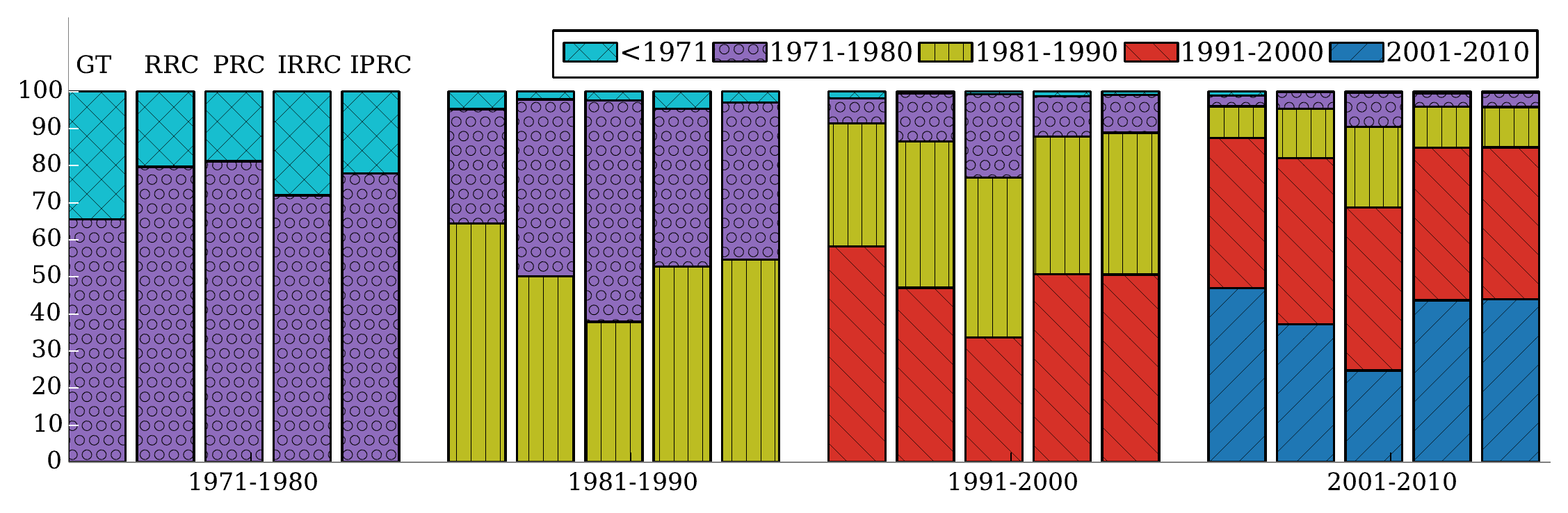}}
  \adjustbox{valign=c}{
  \begin{tabular}{l|r|r|r}
  Simulator ($\lambda,\Theta$) & \rotatebox{90}{Distance} &
  \rotatebox{90}{Turnover} & \rotatebox{90}{Divergence} \\ \hline
  RRC (0.19, 0.9) & 1.08& \textbf{2.70}& 0.03 \\
  PRC (0.3, 0.9) & 1.86& 2.11& 0.16 \\
  IRRC (0.115, 0.8) & 0.60& 2.67& 0.012 \\
  IPRC (0.19, 0.8) & \textbf{0.72}& \textbf{2.70}& \textbf{0.004}
  \end{tabular}}
  \caption{Temporal bucket signatures from ground truth data (GT), random
    relay-cite (RRC), preferential relay-cite (PRC), iterated RRC
    (IRRC) and iterated PRC (IPRC).  $\lambda$ and $\Theta$ were optimized separately for each variant
    using grid search.  Ground truth turnover is 2.70. For
    others, distance, turnover and divergence values are shown in the accompanying table. Note the qualitatively better fit with ground truth compared to Figure~\ref{fig:GT_PA_CP_WYY}.}
  \label{fig:GT_RRC_PRC_IRRC_IPRC}
\end{figure*}

We simulated the model with different values of $(\lambda,\Theta)$.  Grid
search led us to the best value of $(0.19,0.9)$ as per the optimization function defined in ~\eqref{eqn_optimize}.
Figure~\ref{fig:GT_RRC_PRC_IRRC_IPRC} shows the \emph{temporal bucket signatures} for this and the other variants described below; the best $\text{distance}$, $\text{turnover}$ and $\text{divergence}$ that RRC achieves are \textbf{1.08}, \textbf{2.70} and \textbf{0.03}.

\begin{figure}[!htb]
  \begin{mdframed}[backgroundcolor=gray!10]
    \begin{algorithmic}[1]
      \For{advancing time steps $t$}
      \For{each paper $p_n$ newly added at time $t$}
      \For{each citation $(p_n,?)$ to fill}
      \State choose old paper $u$ using PA
      \For{$r = 1, 2, \ldots$}
      \State $T = t-t_u$
      \State toss coin with head prob.\ $\exp(-\lambda T)$
      \State \textbf{if} head or $r > R$: \textbf{break}
      \State toss coin with head prob.\ $\Theta$
      \State \textbf{if} tail: \textbf{break}
      \State let $I(u,t)$ be papers that cite $u$ \\
      \hspace{8em}as of time $t$
      \State choose $v \in I(u, t)$ according to \\
      \hspace{8em}a sampling distribution $D$
      \State $u \leftarrow v$
      \EndFor
      \State add $(p_n,u)$ as new citation
      \EndFor
      \EndFor
      \EndFor
    \end{algorithmic}
  \end{mdframed}
  \caption{Relay-linking template. }
  \label{fig:relay-link-template}
 \end{figure}

\subsubsection{Preferential relay-cite (PRC)}
\label{sec:preferential_relay-cite_model}

In the preferential relay-cite model, $R$ continues to be 1, but we depart
from the random relay-cite model in that $D$ is no more a uniform
distribution over the papers in $I(u,t)$.  The probability of sampling
$v$ is proportional to its in-degree, as in PA.  Again, we simulated
this model and performed a grid search to obtain the best parameter values $(\lambda,\Theta) = (0.3, 0.9)$ as per the optimization function in Equation~\ref{eqn_optimize}. We obtained the best $\text{distance}$ score of
\textbf{1.86}. The corresponding $\text{turnover}$ and $\text{divergence}$ scores were found to be \textbf{2.11} and \textbf{0.16}.

\subsubsection{Iterated random relay-cite (IRRC)}
\label{sec:iterated-relay-cite}
In iterated random relay-cite model, we relax $R$ to be able to follow
the relay-cite hypothesis iteratively. Thus, once a paper $v$ has
sampled a paper from $I(u, t)$ based on uniform distribution, 
we again toss a coin with head probability = $\exp(-\lambda T')$, where $T'$ is the current age of the paper $v$. In case, tail turns up, we follow this
process recursively. $(\lambda,\Theta)=(0.115,0.8)$ gives the
best $\text{distance}$ score of \textbf{0.60}, $\text{turnover}$ of \textbf{2.67} and $\text{divergence}$ score of \textbf{0.012}.

\subsubsection{Iterated preferential relay-cite (IPRC)}
\label{sec:Recursive_Preferential_Delinking_Model}
In iterated preferential relay-cite model, once a paper $v$ has sampled
a paper from $I(u,t)$ based on PA, we again toss a coin with head probability = $\exp(-\lambda T')$, where $T'$ is the current age of the paper $v$. In case, tail turns up,  we follow this process recursively. We
simulated the model with different parameter values, and found that $\lambda
= 0.19$ and $\Theta = 0.8$ gives the best $\text{distance}$ score of \textbf{0.72}, $\text{turnover}$ score of \textbf{2.70} and $\text{divergence}$ score of \textbf{0.004}.

\subsection{Dependence on bucket size}
Since $\text{divergence}$ is computed from age gap count histograms,
it does not depend on the bucket size. For $\text{distance}$ and $\text{turnover}$, 
we observed that our observations are stable for bucket sizes 7, 8 and 9 years.
For bucket sizes larger than 10 years, the number of buckets is too small to make a fair comparison.

\section{Comparison between models}
\label{relay_main_results_good_fit}

\subsection{Temporal bucket signatures}
Fig.~\ref{fig:GT_RRC_PRC_IRRC_IPRC} compares ground truth (GT)
temporal bucket signatures against the variations of relay-linking
models described above.  Three out of four relay-linking models proposed above
outperform the popular baseline models of network evolution
in terms of all the observables, i.e., $\text{distance}$, $\text{turnover}$ and $\text{divergence}$ (see Figure~\ref{fig:GT_PA_CP_WYY} for detailed result obtained for the
baseline models.)  Further, note that IPRC outperforms all the other relay-linking models in at least two out of the three observables and can be considered to be the closest fit to GT.
Therefore, in order to strengthen our results, we compare age gap count histograms and degree distribution of IPRC (instead of other relay-linking models) with the baseline models.

\subsection{Age gap count histograms}
\begin{figure}[!tbh]
\centering\includegraphics[width=.8\hsize]{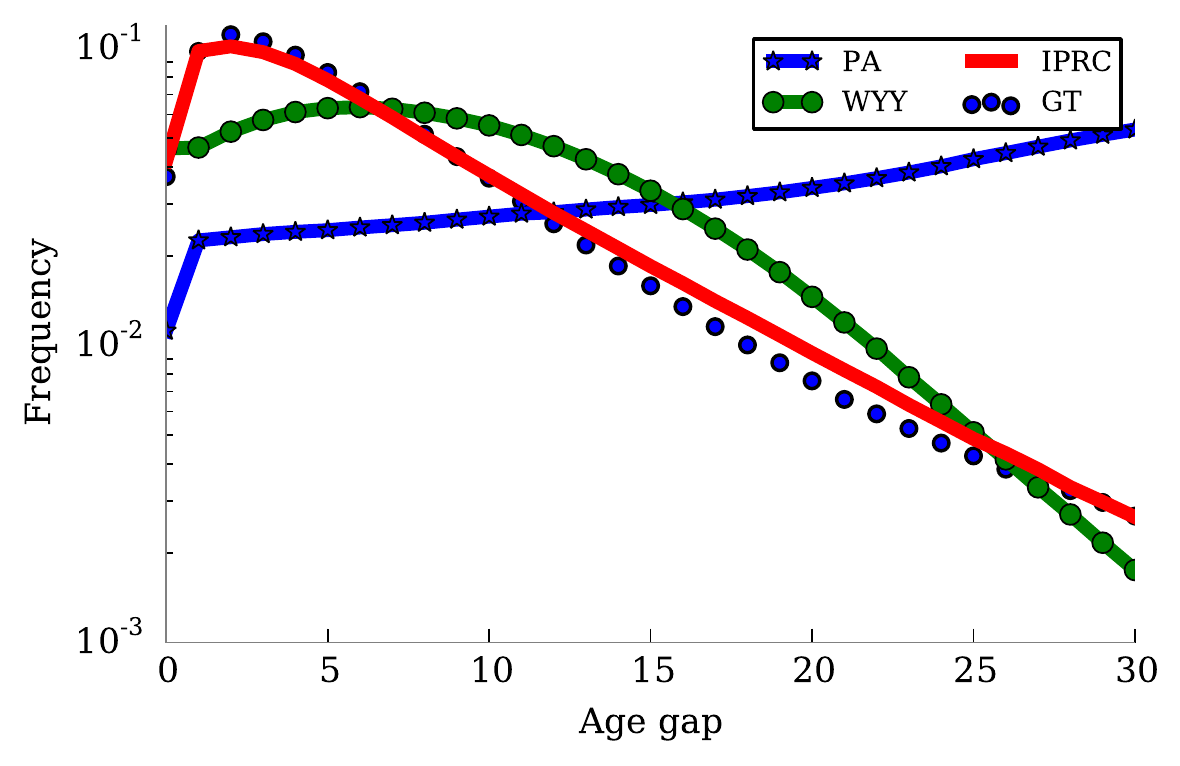}
  \caption{Age gap count histograms. WYY is quite close to
    ground-truth, but for its best choice of $\lambda$, its peak is still at too large a gap. IPRC's decay fits GT best. The $\text{divergence}$ values are, PA: 0.77; WYY($\lambda = 0.11$): 0.13; IPRC ($\lambda = 0.19,\Theta = 0.8$): 0.004}
  \label{tab:Expt:AgeGapCount}
\end{figure}

Fig.~\ref{tab:Expt:AgeGapCount} shows the age gap count histograms
defined in equation \eqref{eq:AgeGapCount} for various simulators, compared with ground
truth (over all time).  Ground truth rolls down steadily after an early peak at 2--3 years age gap. As expected, the PA curve keeps going up, because aging is always an advantage. Surprisingly, but indirectly corroborating degree distribution (as well as its temporal signature in Figure~\ref{fig:GT_PA_CP_WYY}), WYY does well in comparison, but its most likely gap is larger compared to real data.
IPRC fits GT's decay best.

The model complexity of
relay-linking is comparable to PA.  Yet, we
establish that relay-linking is the closest to real networks in terms
of $\text{divergence}$, $\text{distance}$, and $\text{turnover}$.

\subsection{Degree distribution}
\label{sec:recheck-degree}

\begin{figure}[!htb]
  \centering
  \includegraphics{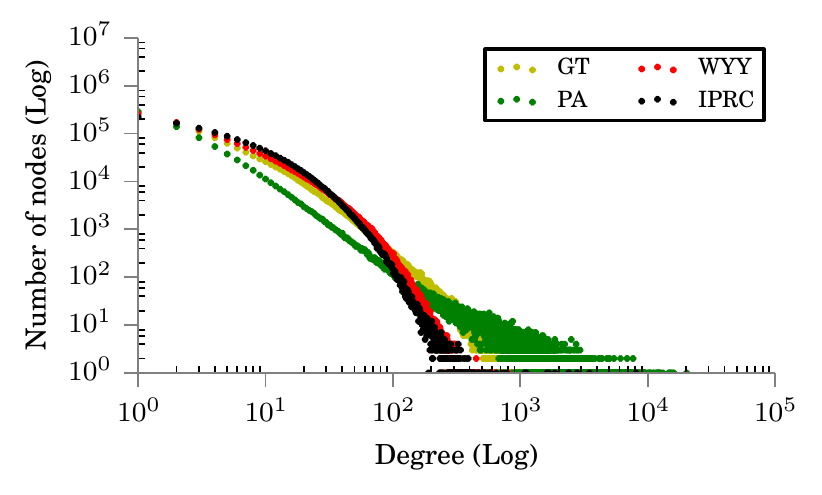}
  \caption{Degree distributions of ground truth (GT) and various
    models (PA,WYY,IPRC) at the best optimal parameters values.}
  \label{fig:recheck-degree}
\end{figure}

We also find it remarkable that relay-linking models fit temporal bucket signatures better
than all other models.  In Figure~\ref{fig:recheck-degree} we plot the degree distribution of the network obtained by simulating IPRC. The figure shows that the distribution fits the GT quite well. We should, however, verify that other properties of real networks that are matched well by preferential attachment or similar models are preserved.

\section{Practical Application}
\label{sec:TurnoverImpactFactor}
To get more insight into temporal bucket signatures, we apply these to a cross-sectional study by sub-field and conference slices.
The widely quoted \emph{impact factor}~\cite{garfield2006history}
(IF10) of a journal or conference is the average number of citations
to recent (last 10 years) articles published there.
Table~\ref{decay_impact_factor} shows the turnover values we estimate
against IF10 for the four conference subsets we chose.  There is a
clear negative correlation i.e., communities with large turnover
have low IF10.  Large turnover also seems associated with applied
communities in a state of more intense flux.

\begin{table}[!tbh]
 \centering 
  \caption{Correlation between turnover and average value of 10-year
    impact factor, over specific conferences as well as coherent
    sub-communities of computer science. Note the negative correlation between turnover and 10-year
    impact factor. Communities with large turnover have low IF10.}
  \label{decay_impact_factor}
  \begin{tabular}{|p{.55\hsize}|c|c|}
  \hline
  \textbf{Conference Name}&\textbf{Turnover}&\textbf{Avg. IF10}   \\\hline
  SIGMOD&3.97&3.50\\\hline
  VLDB,ICDE&4.52&2.79\\\hline
  SIGIR&5.61&2.77\\\hline
  ICML,NIPS&6.74&1.84\\ \hline \hline
  Data Mining, machine
learning, artificial
intelligence, natural language
processing and information
retrieval &3.32&0.63\\  \hline
Distributed and parallel
computing, hardware and
architecture, real time and
embedded systems & 3.31 & 0.74 \\ \hline
Algorithms and Theory,
Programming Languages and
Software Engineering & 2.29 & 0.78\\ \hline
 \end{tabular}
\end{table}

\section{Conclusion}
\label{sec:end}

Idealized network evolution models that explain entrenchment of
prominence are abundant, but the only ones that model aging depend on
post-hoc distribution-fitting (data collapse) and externality
(fitness) parameters.  We give the first plausible network-driven
models for obsolescence in the context of research paper citations,
based on a natural notion of \emph{relay-linking}.  Studying large
bibliographic data sets, we also propose several novel and stringent
tests for temporal fidelity of evolving, aging network models.
Traditional aging models do not pass these tests well, but our
relay-linking models do.

Finally, a number of potential limitations need to be considered. First, the current study employs bibliographic datasets only. Therefore, we do not claim about generic applicability in other social networks. In future, we plan to extend this study to other citation networks, for example, U.S. Supreme Court citation network. Second, our proposed relay models do not consider area/author information which might be relevant in deciding the relay citation.
On account of the fact that the current work is only a preliminary attempt to understand the relaying phenomenon of citation links, future extensions could possibly lead to formal analysis of properties of relay-linking or tractable variations.

\bibliographystyle{ACM-Reference-Format}
\bibliography{sigproc}
\flushend
\end{document}